# Cavity-Mediated Radiative Energy Transfer Enables Stable, Low-Threshold Lasing in Hybrid Quantum Dot-Nanoplatelet Supraparticles

Cristian Gonzalez[1], Yun Chang Choi[1], Gary Chen[1], Jun Xu[3], Claire Yejin Kang[1,2], Emanuele Marino[4]*, Cherie R. Kagan[1,2,3], Christopher B. Murray[1,2]*

[1] Department of Chemistry, [2] Department of Material Science & Engineering, and [3] Department of Electrical & System Engineering, University of Pennsylvania, Philadelphia (PA) 19104, USA.

[4] Dipartimento di Fisica e Chimica – Emilio Segrè, University of Palermo, Palermo 90123, Italy

*Corresponding author: cbmurray@sas.upenn.edu, emanuele.marino@unipa.it

ORCID:

Cristian Gonzalez: 0000-0001-7639-6193

Yun Chang Choi: XXXX-XXXX-XXXX-XXXX

Gary Chen: 0009-0001-3503-1654

Jun Xu: 0000-0003-4617-6602

Claire Yejin Kang: 0009-0006-3982-1748

Emanuele Marino: 0000-0002-0793-9796

Cherie R. Kagan: 0000-0001-6540-2009

Christopher B. Murray: 0000-0002-3491-122X

Abstract

Colloidal semiconductor nanocrystals are promising building blocks for optoelectronics due to their solution processability, spectral tunability, and ability to self-assemble into complex architectures. However, their use in lasing application remains limited by high working thresholds, rapid nonradiative losses from Auger recombination, and sensitivity to environmental conditions. Here, we report hybrid microscale supraparticles composed of core/shell CdSe/ZnS quantum dots (QDs) and CdSe/$Cd_xZn_{1-x}$S nanoplatelets (NPLs), which overcome these limitations through efficient, cavity-mediated energy funneling and coupling. Broadband absorbing QDs rapidly transfer excitation to narrow emitting NPLs, enabling stable whispering gallery mode lasing with a low threshold of 0.35 mJ/cm². These supraparticles retain optical performance after prolonged exposure to air, water, and continuous irradiation, offering practical advantages for optoelectronic devices and advanced pigment technologies. Ultimately, our approach provides a versatile, programmable platform for optical amplification and tunable emission control within colloidal photonic architectures.

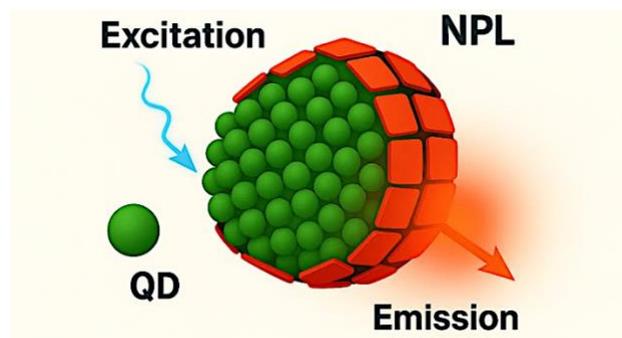

Introduction

Colloidal semiconductor nanocrystals (NCs) have attracted intense interest as versatile optical materials for both active photonic devices and functional pigments.[1, 2] Isotropic quantum dots (QDs) exhibit 3D quantum confinement, resulting in size-tunable, narrow emission characterized by near-unity photoluminescence quantum yields, valuable for lasers, displays, and luminescent coatings.[3, 4] Despite these advantages, the spatial confinement of charge carriers facilitates rapid nonradiative Auger recombination, causing undesirable optical losses. Compositionally graded nanostructures have been shown to limit these losses, but more general strategies to avert Auger recombination would significantly improve QD brightness and optical gain.[5] One strategy consists in limiting carrier confinement along certain directions.[6]

Colloidal nanoplatelets (NPLs) feature 2D quantum confinement, larger exciton oscillator strengths, and faster radiative recombination rates relative to QDs.[7] Through their reduced spatial confinement and in-plane delocalization of charge carriers, NPLs exhibit lower Auger recombination, resulting in extended multiexciton lifetimes.[8, 9] These characteristics result in lower lasing thresholds,[10] making NPLs valuable as stable, non-fading optical pigments for high-tech coatings. While QDs offer practical advantages such as easier synthesis, surface modification, and optical tunability across a broader spectral range, NPLs feature narrow emission defined by their atomically precise thickness.[11-14] The targeted combination of QDs and NPLs may result in superior optical properties by harnessing the benefits of both systems while offsetting their limitations. Such a profitable combination may be achieved through self-assembly.

Self-assembly is a well-established, nature-inspired tool to organize NCs into functional superstructures,[15-20] providing additional avenues for both photonic and pigment applications.[21-23] For instance, the emulsion-templated assembly of NCs leads to spherical supraparticles (SPs) with remarkable optical properties.[24-26] NC SPs feature collective photonic resonant modes with high quality factors (Q), such as Mie[27, 28] or whispering-gallery modes (WGMs),[29, 30] that enable enhanced emission rates and lasing.[31, 32] Overall, each SP functions as an independent, self-contained optical unit that integrates a gain medium and a resonator. This design eliminates the need for external, substrate-bound resonators that impose material and geometric constraints and that are distinguished from the medium, such as etched waveguides or distributed Bragg reflectors.[32, 33] Instead, SPs behave as optically active colloids, readily deployable and applicable to both solution-processable photonics and functional coatings.[34, 35] These coatings may be realized through the assembly of these SPs into higher-order structures, leading to the emergence of hierarchical physical properties, such as structural color.[22, 36]

Despite their promise as multifunctional materials for lasers and optical pigments, SP systems based on QDs face several challenges. Firstly, the high lasing thresholds observed arise primarily from high Auger recombination in the QDs, which limits multiexciton lifetimes and requires large energy inputs to achieve carrier population inversion. While shell growth can mitigate Auger effects, especially in thick shell "giant" QDs, many core/shell designs still exhibit non-negligible Auger losses. Secondly, the optical performance of SPs suffers from reversible and irreversible spectral instabilities due to photoinduced refractive index changes under high excitation fluences, or by absorption-limited mode selection,[37, 38] leading to undesirable spectral drift and tunability of lasing emission[39]. For lasing applications, such instability jeopardizes coherent emission; for pigment or coating applications, it compromises reproducibility of spectral features vital for use cases such as anti-counterfeiting, where narrowband signatures must remain consistent.[40-43] Bulk lasers overcome these instabilities through protective encapsulation, while SP-based materials must retain optical integrity under exposure to air, moisture, and intense light to capitalize on their suitability for deployment to complex environments. Therefore, an effective approach must pair materials with reduced nonradiative losses with an environmentally resilient resonator architecture to enable stable, low threshold lasing suitable for free-standing, robust optical materials.

We exploit the optical cooperation between two core/shell NC species, CdSe/ZnS QDs and CdSe/Cd$_x$Zn$_{1-x}$S NPLs, by their co-assembly within the same SP, resulting in compact cavity-coupled microsystems optimized for stable, low-threshold lasing. Using a source–sink emulsion microfluidic process, we fabricate SPs at an estimated throughput of 28,800 SPs/h. The QDs allow for broadband absorption of light from the visible to the ultraviolet thanks to their small CdSe cores and thick ZnS shell, making them ideal energy donors. Meanwhile, NPLs with high oscillator strengths and fast radiative recombination function as efficient lasing centers. Within the SP, these two components form a cavity-mediated radiative energy transfer pathway, enabling efficient excitation funneling from QDs to NPLs. Time-resolved photoluminescence (TRPL) measurements on hybrid QD-NPL SPs, revealing a ~91 % reduction in emission lifetimes, consistent with strong coupling between emitter and cavity. This enhancement, together with efficient radiative energy transfer from QDs to NPLs, enables selective population inversion in the NPLs, resulting in stable lasing with a low threshold of 0.35 mJ/cm², significantly lower than that of SP lasers based only on QDs or NPLs.[37]

In addition to their favorable photo physics, lasing QD-NPL SPs show minimal spectral drift of <0.8 nm over 20 minutes of operation. These hybrid SPs maintain consistent optical performance after prolonged exposure to ambient air in high excitation fluences, underscoring their environmental durability and suitability for sensors, coatings, and photonic pigments.

Results and Discussion

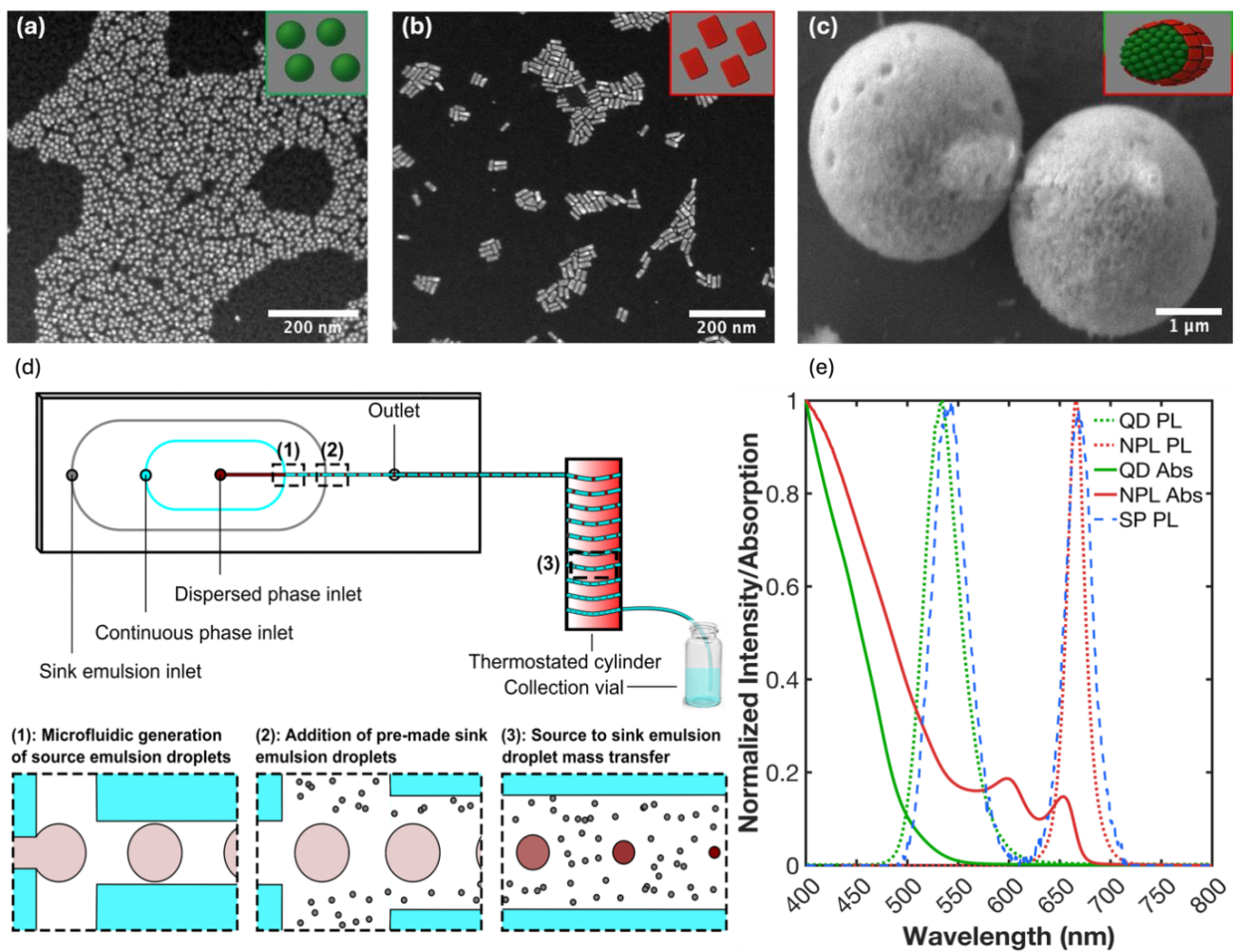

Figure 1: Synthesis and assembly of hybrid SPs composed of CdSe/ZnS QDs and CdSe/Cd$_x$Zn$_{1-x}$S NPLs. (a,b) STEM images of QDs spherical morphology (9.1 nm $\pm$ 0.6 nm) and NPLs anisotropic geometry (28.3 nm $\pm$2.7 nm $\times$ 8.9 nm $\pm$1.8 nm) before assembly. (c) SEM image of a representative hybrid SP formed via droplet microfluidics, showing spherical assemblies (3.56

µm ± 0.22) and subtle surface features indicative of exposed NPLs. (d) Schematic of the source-sink droplet microfluidic assembly method. QDs and NPLs dispersed in toluene are emulsified in SDS-stabilized aqueous phase (source), then mixed with a secondary SDS-stabilized emulsion containing hexadecane in water (sink). Toluene diffusion drives NC concentration and densification into SPs. (e) Normalized optical absorption and photoluminescence spectra of individual components and the resulting hybrid 50:50 SP.

Figure 1 outlines the synthesis and assembly of hybrid SPs composed of colloidal CdSe/ZnS QDs and CdSe/Cd$_x$Zn$_{1-x}$S NPLs. Figures 1 (a, b) show scanning transmission electron micrographs (STEM) of QDs and NPLs before assembly. The QDs exhibit a near spherical morphology with a diameter of 9.1 nm ± 0.6 nm. The CdSe core diameter was estimated by using empirical sizing curves and the PL maximum, assuming negligible spectral shift upon ZnS overcoating due to the large bandgap of ZnS and strong carrier confinement. This yields an estimated core size of 3.6 nm and a 2.75 nm corresponding shell thickness per side.[44] The NPLs display planar shapes, with a long axis of 28.3 nm ± 2.7 nm and a short axis of 8.9 nm ± 1.8 nm consisting of 4.5 monolayer CdSe cores passivated with a thick, graded Cd$_x$Zn$_{1-x}$S shell. Figure 1 (d) features the droplet microfluidic approach used to assemble the SPs.[45] A mixture of QDs and NPLs in toluene (1:1 by volume, 0.01% initial volume fraction) is introduced to a glass microfluidic chip filled with an aqueous solution of sodium dodecyl sulfate (SDS, 20 mM). We adjust the input pressures of the inlets to reach a stable dripping regime that produces monodisperse droplets at a rate of ~28,800 SPs/h, as analyzed from Figure S1. SDS acts as a surfactant that stabilizes the monodisperse emulsion referred hereon as the "source" emulsion (Step 1). Separately, we prepare a "sink" emulsion via ultrasonication consisting of 1% hexadecane in water (SDS, 20 mM). This emulsion is introduced into the microfluidic chip through an additional inlet (Step 2). Mixing source and sink emulsions initiates a unidirectional mass transfer of toluene, concentrating the NCs within the source droplets and swelling the sink droplets (Step 3). We vary the co-residence time and temperature of source and sink emulsions to allow for the formation of densely packed, monodisperse SPs. Figure 1(c) presents a scanning electron micrograph (SEM) of the resulting hybrid SPs, yielding spherical super assemblies roughly 3.56 µm ± 0.22 µm. While SEM provides the overall morphology, it does not resolve the finer surface details. To probe these features, we examined smaller, microemulsion-formed SPs by HRTEM, shown in Figure S2, which revealed surface texture with elongated features resembling exposed NPLs. This observation suggests a core-shell spatial arrangement, where the QD component preferentially occupies the core while the NPLs localize at the surface. This is consistent with a previous study from our group where larger NaGdF$_4$ disks occupied the shell of the SPs while smaller PbS/CdS QDs localized to the core,[45] as well as other studies displaying preferential separation within droplets.[46, 47] This phase separation behavior is driven by thermodynamic considerations, where shape differences, packing constraints, and surface energy guide the QDs and NPLs into energetically favorable domains during the densification of the native droplets.[48] While limited intermixing cannot be fully excluded, our structural and photophysical characterization is consistent with the dominance of NPLs at the SP surface.

Figure 1 (e) displays the optical absorption and photoluminescence (PL) for both QD and NPL components and their hybrid SP. The QD emission spectrum peaks at ~533 nm, while the NPLs peaks at ~667 nm. The QD emission overlaps completely with the NPL absorption, suggesting ideal conditions for efficient radiative energy transfer from QDs (donor) to NPLs (acceptor). The

PL displayed from a single hybrid SP shows the emission of both QDs and NPLs, with slight red shifts of ~7 nm and ~3 nm respectively, likely related to changes in the local dielectric environment and coupling effects within the densely packed structure.[49]

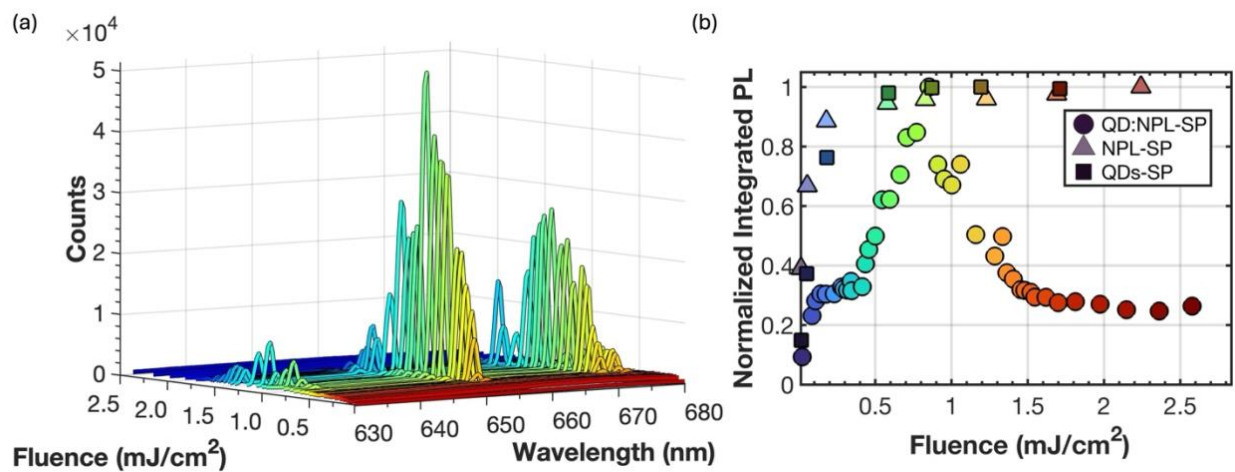

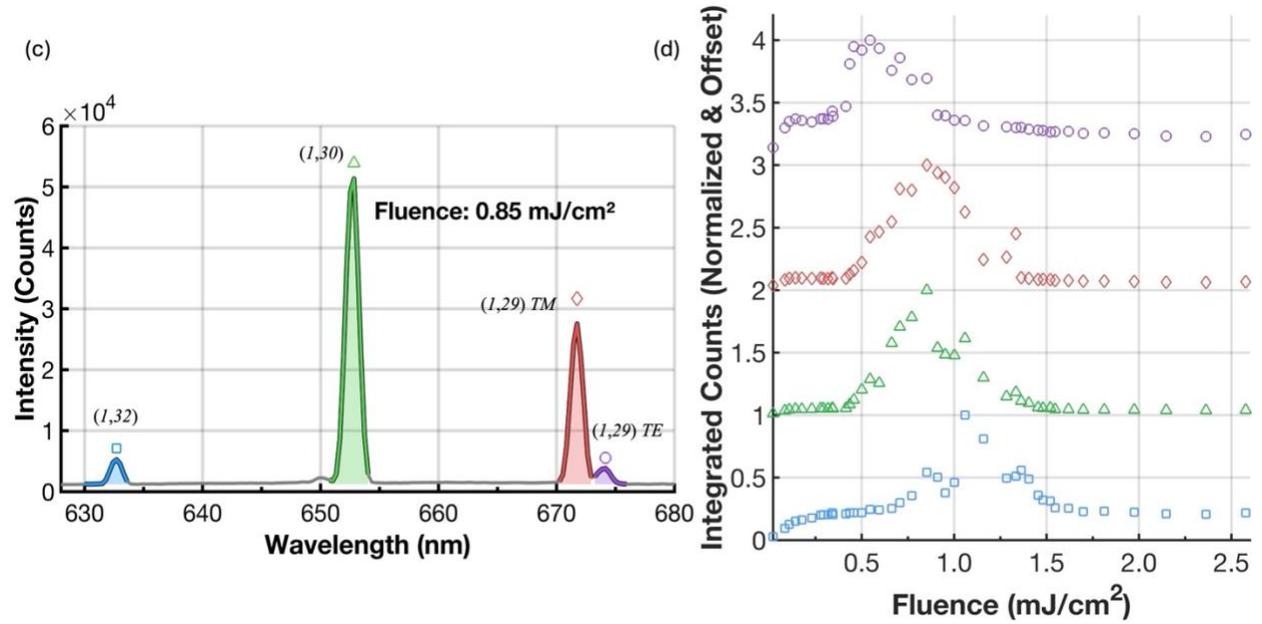

Figure 2: Optical characterization and lasing behavior of hybrid SPs under 488 nm excitation. (a) Evolution of emission spectra with increasing excitation fluence for hybrid QD: NPL-SPs, and (b) corresponding lasing curves for hybrid and individual component SPs (QDs-SP and NPL-SP), traces are color-coded from red to blue with increasing excitation fluence.(c) Detailed lasing spectrum at 0.85 mJ/cm$^2$, showing fundamental radial whispering-gallery modes (WGMs) labeled by radial and azimuthal mode indices (n,*l*). Color-coded regions denote spectral windows for fluence-dependent intensity analysis. (d) Mode-dependent lasing thresholds and relative integrated intensities as a function of fluence.

We study the lasing characteristics of individual SPs by using the optical setup shown in Figure S4. Excitation at 488 nm was provided by a pulsed laser source (272 fs pulse width, 10 kHz repetition rate), coupled to an optical parametric amplifier for wavelength tuning. The output from this system is guided through an optical fiber to a confocal microscope, with an objective to allow single SP excitation. The emission from a single SP is collected and directed through optical fibers either to a monochromator and a silicon CCD for spectral measurements or to a time-correlated single-photon counting detector for lifetime measurements. The fluence is varied using neutral density filters to enable fluence-dependent spectral measurements. Figure 2 (a) illustrates the evolution of the emission spectra of a single hybrid QD:NPL-SP with increasing excitation fluence. Between 0.02 mJ/cm$^2$ and 0.35 mJ/cm$^2$ we observe a PL signal like that shown in Figure 1(e), highlighting the presence of both QDs and NPLs within the SP. The spectral window of Figure 2 (a) (630–680 nm) was chosen to resolve the NPL WGM features, so the QD PL band at 540–560 nm falls outside this range and is shown in Figures S7. Around 0.35 mJ/cm$^2$, four distinct, sharp peaks emerge from the broader emission envelope of the NPLs. Between 0.35 mJ/cm$^2$ and 0.85 mJ/cm$^2$ we observe a narrowing in the average spectral width of the peaks from ~2.4 nm to ~1.2 nm, as well as a 40-fold increase in their intensity. This nonlinear behavior implies the onset of population inversion and marks the transition from incoherent spontaneous emission to coherent stimulated emission, indicative of lasing. No lasing peaks were observed within the emission envelope of the QDs. We designate "lasing" because thresholded, mode-resolved WGM peaks with cavity-set spacing and mode competition are observed; ASE would yield a single narrowing band without discrete resonances. The broader linewidths reflect the finite Q of the supraparticles, not ASE.

We evaluate the stability and robustness of the hybrid SP under high excitation conditions. The SP reaches its maximum recorded intensity around 0.85 mJ/cm². This suggests that between 0.35 mJ/cm² and 0.85 mJ/cm² the optical gain induced by increased excitation fluence surpasses losses from nonradiative Auger recombination. Beyond 0.85 mJ/cm², the intensity of the lasing peaks diminishes. We propose that in this regime nonradiative losses, particularly Auger recombination, begin to dominate over optical gain. This behavior continues until 1.54 mJ/cm², where the distinct lasing peaks completely collapse back into the broader emission envelope of the NPLs. Accounting for fluence, spot size, absorption cross-section, and assuming partial excitation due to a ~1 μm light penetration depth in the SP at 488 nm excitation (Justification in SI Section 1.0), we estimate ~2.7 × 10⁶ NCs are effectively excited within the SP (~10⁷ NC total per SP), yielding average exciton populations of ~ 2.1 excitons per NC at 0.85 mJ/cm² and ~ 6.3 excitons per NC at 2.58 mJ/cm² (See SI for full derivation). These values place the system deep within the multiexciton regime where Auger recombination is expected to dominate.[50] Further increases beyond 1.54 mJ/cm² result exclusively in broad NPL emission with relatively constant counts (~5.0 × 10⁴), demonstrating negligible enhancement of the radiative emission processes up to the highest tested fluence of 2.58 mJ/cm². After exposure to higher fluences, we observe the recovery of the lasing behavior when the excitation fluence is reduced to 0.85 mJ/cm², as shown in Figure S11. In particular, the lasing peaks show nearly identical intensities and minimal spectral blue shifts (~2.0 nm), except for the highest energy mode at 632.6 nm, which does not recover. We attribute these observations to minor refractive index changes and mode competition effects, where lower energy-modes with lower lasing thresholds consume the available gain first. This gain saturation can inhibit the reappearance of higher-energy modes reaching threshold again. Figure 2 (b) shows the normalized integrated intensity of the emission spectrum as a function of excitation fluence for the hybrid QD:NPL-SP, capturing in a compact format the results of Figure 2 (a).

We compare these results for hybrid QD:NPL-SPs to those obtained for individual SPs obtained from the assembly of only QDs (QD-SP) and NPLs (NPL-SP), showing no sign of amplification. Both samples exhibit linear increases in PL counts with fluence, characteristic of spontaneous emission. The QD-SP (3.62 μm ± 0.17 μm) and NPL-SP (3.40 μm ± 0.12 μm) are similar in size, as shown in Figure S12, ruling out geometric factors as the cause of the absence of lasing. The QD-SPs saturate at approximately 0.58 mJ/cm², while the NPL-SP show a shallower linear increase at higher fluences. The absence of amplified emission in either single-component SP highlights the critical role of the synergistic design of the hybrid SP.

Figure 2 (c) shows the lasing spectrum obtained from the hybrid QD:NPL-SP at 0.85 mJ/cm² excitation fluence. We associate the four dominant lasing peaks with four whispering gallery mode (WGM) resonances that can be identified by a pair of radial and azimuthal mode numbers (*n,l*). To assign these, we used the approximate WGM resonance condition for dielectric microspheres:

$$\lambda_{n,l,P} \approx \frac{2\pi \mathcal{R} n_{eff}}{l + f(n,P)}$$

where $\mathcal{R} \approx 1.78 \mu m$ is the SP radius determined by SEM, and $n_{eff}$ is treated as the sole fitting parameter. We varied $n_{eff}$ until the predicted and observed peak positions matched, which yielded $n_{eff}$ =1.83. This fit uniquely assigns the 632.6 nm mode as (1,32), the mode at 652.6 nm as (1,30), and the two closely spaced modes at 671.7 nm and 674.2 nm as the TM and TE components of the (1,29) mode, respectively.

Finite-difference time-domain (FDTD) simulations further support our model, revealing two distinct resonances in this region: one broader, weaker mode, and another sharper and more confined mode consistent with TE and TM polarization-dependent field confinement. While definitive assignment requires polarization-resolved measurements, the strong agreement between theory, simulation, and experiment suggests that this doublet originates from intrinsic TE-TM splitting of a nearly degenerate WGM, likely enhanced by broken spherical symmetry due to the substrate, surface perturbations, and structural asymmetries in the SP. (Figure S13) While WGMs are highly sensitive to cavity size, our microfluidic source–sink emulsion method yields supraparticles with < 2 % size variation,[45] sufficient to achieve reproducible mode positions and thresholds. For a ~3.5 µm SP, a ± 2 % diameter variation corresponds to a fractional mode shift $\Delta\lambda/\lambda \approx \Delta D/D$; at $\lambda \approx 650$ nm this amounts to $\approx \pm 13$ nm (and $\approx 6–7$ nm for ± 1 %), demonstrating practical wavelength control within predictable fabrication tolerances.[51, 52] Furthermore, WGM resonances can be fine-tuned post-fabrication by small refractive-index changes or mild thermal adjustments, providing a secondary control handle at the 1–5 nm level.

Efficient outcoupling from WGM resonators is routinely achieved using evanescent-field coupling schemes, such as tapered fibers, prisms, or integrated waveguides, all of which are directly compatible with our SP geometry.[32, 53-56]

The spectral width $\Delta\lambda$ of a lasing peak reflects the quality of the resonator, namely, how efficient the corresponding photonic mode is at trapping light. For each mode, we calculate an active (hot-cavity) quality factor extracted from the lasing peaks above threshold, defined by Q = $\lambda/\Delta\lambda$, such that sharper peaks indicate more efficient trapping. At the peak performance fluence of 0.85 mJ/cm², the extracted Q-factor values are 537 (632.6 nm), 454 (652.6 nm), 539 (671.7 nm), 548 (674.2 nm) listed in order of increasing wavelength (Figure S14), suggesting that light remains trapped in each WGM at the surface of the SP for a lifetime $\tau = \frac{Q\lambda}{2\pi c}$ between 157 and 196 fs. This duration implies an effective photon travel path $L_e = \frac{c}{n}\tau$ ranging from ~26 µm-32 µm, for a refractive index of $n \approx 1.83$. Notably, this is equivalent to roughly 2-3 circulations around the SP circumference $C = 2\pi\mathcal{R} \approx 11\,\mu m$, confirming that lasing modes persist through multiple complete WGM round-trips. These Q values are lower than those reported from our previous works with Q = 1900 ± 400, likely because of the smaller size of the SPs and the larger size of the NPLs at the surface of the SPs, contributing to optical losses through scattering.

The estimated gain bandwidth, defined by the full width at half maximum (FWHM) of the NPL PL, is approximately 20.8 nm (655.4 nm to 676.2 nm), with a broader estimation of ~41.6 nm if considering the spectral distance between the highest (674.2 nm) and lowest (632.6 nm) observed lasing peaks.

To understand how the gain bandwidth interacts with cavity modes, we analyze the dependence of the integrated intensity of each lasing peak on excitation fluence after subtracting the background emission envelope in Figure 2 (d), Interestingly, we observe that the fluence at which each mode (n,l) reaches its peak intensity depends on l, indicating mode-dependent amplification behavior. The mode at 632.6 nm ($l$ = 32, blue squares) reaches maximum intensity at a higher fluence of 1.02 mJ/cm$^2$. Modes at 652.6 nm ($l$ = 30, green triangles) and 671.7 nm ($l$ = 29, red diamonds) peak at 0.85 mJ/cm$^2$, while the mode at 674.2 nm ($l$ = 29, purple circle) peaks at an even lower fluence of 0.51mJ/cm$^2$. Because the estimated gain peaks near 667 nm, the short-wavelength mode at 632.6 nm sits ~ 34 nm at the edge of the gain window, whereas the long-wavelength mode at 674.2 nm is only ~7 nm away.

We attribute this asymmetric gain behavior to the interplay between two competing factors: (1) the detuning between the spectral position of the mode and the gain peak and (2) the intrinsic cavity losses. While modes with higher $l$ values possess higher intrinsic Q-factors due to tighter confinement of the optical field near the SP surface, they may appear at the spectral edges of the gain window, where amplification is limited.[57] This spectral mismatch results in over a 90% decrease in peak intensity compared to centrally aligned modes, thereby requiring higher excitation fluences. This is consistent with WGM theory, where increasing $l$ leads to reduced radiative leakage and longer optical path lengths, and thus enhanced mode quality and minimized losses.[55, 58]

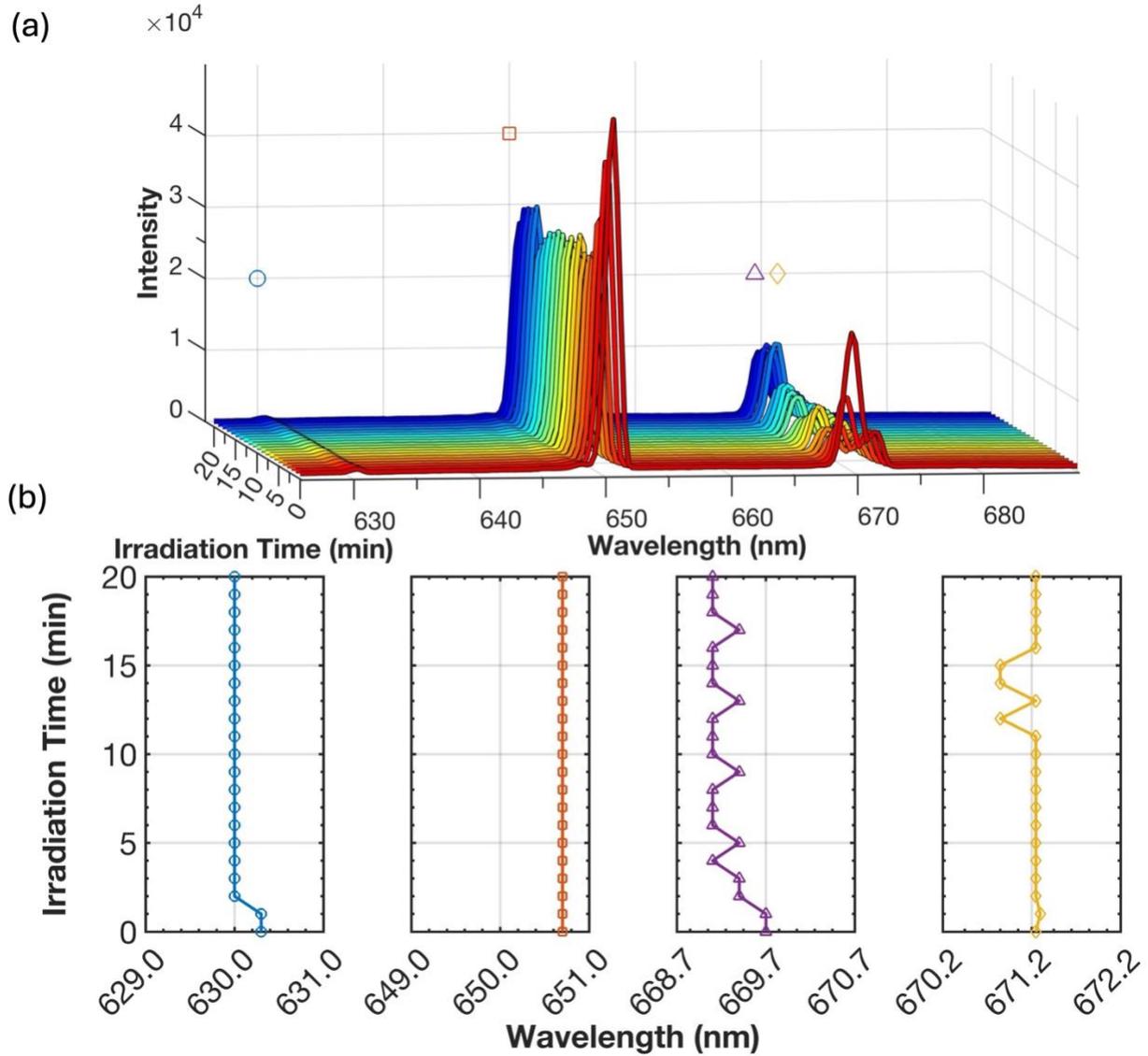

Figure 3: Long-term spectral stability and lasing performance of hybrid QD-NPL SP. (a) continuous lasing emission monitored over 20 min under constant excitation at ~1 mJ/cm², nearly a factor of three above the lasing threshold. (b) close examination of spectral positions of four prominent lasing modes during continuous excitation.

In practical applications, including microlasers and photonic pigments, stability is a crucial requirement for commercialization. Figure 3 highlights the stability of the laser emission of our hybrid QD-NPL SPs. As shown in Figure 3(a), under an excitation fluence of approximately 1 mJ/cm², nearly a factor of 3 higher than the threshold of 0.35 mJ/cm², a single SP maintains lasing behavior under pulsed excitation at 10 kHz over 20 min of observation. We observe no appreciable shifts in the two higher energy modes and oscillations within 0.8 nm for the two lower energy peaks. These time-dependent spectra were then used to extract peak position traces shown in Figure 3(b), where the local maximum of each lasing mode was tracked over time. The resulting averages and standard deviations of the spectral maxima of each lasing mode are 630.15 $\pm$ 0.15 nm (blue), 650.7. ± ≤ 0.05 nm (orange), 669.7 ± 0.4 nm (purple), and 671.25 ± 0.35 nm (gold), revealing that the lowest energy modes experience an initial ~ 0.3 nm blueshift during the first two minutes, after which all four modes settle within ± 0.15 nm for the remainder of the laser operation. To further quantify this stability, we have added time-resolved analyses of peak intensity and FWHM for each lasing mode as well as 2D projections (See SI, Figure S15, S18), showing that all four peaks remain cavity-narrow and stable after initial equilibration. In previously reported CQD SP microlasers, lasing modes typically blue-shift by >30 meV (≈10–15 nm) within ~15 min under constant pumping, whereas here all four WGMs remain within $|\Delta\lambda| \leq 0.15$ nm for ≥ 20 min. Spectral instabilities in lasing systems have been linked to refractive index changes via the optical Kerr effect and modulational mode shifts.[30, 56, 59, 60] The minimal drift observed here is consistent with a cavity that suppresses those mechanisms Although rigid microcavities can achieve sub-pm stability, our ≤ 0.15 nm drift over ≥ 20 min corresponds to a >×10 reduction in spectral drift relative to prior CQD SPs under similar constant-pump tests,[37] and is comparable in duration, but far tighter in wavelength stability than solidified CQD microlasers that report ~40 min operational stability at 450 K without sub-nm spectral locking.[37, 61] We hypothesize that the slightly larger excursions of the longer wavelength modes arise from their lower lasing thresholds, which leave them more sensitive to transient carrier density fluctuations and surface index changes; subtle cavity-mode competition may further amplify these early time shifts. Once steady-state gain is reached, carrier redistribution damps these effects, locking modes in place.

Two design features likely underpin this robustness: (1) the 2-D geometry and in-plane excitonic dipoles of the NPL, which reduce out-of-plane local field coupling and thereby mitigate Kerr-type index variations, [62, 63] and (2) continuous QD-to-NPL energy transfer that serves as an auxiliary pump, smoothing carrier densities and sustaining optical gain over extended operation. Collectively, these factors yield a hybrid SP laser whose emission remains locked to within a few tenths of a nm for at least 20 minutes. This level of ambient-air stability, sustained lasing over >$10^7$ excitation cycles without encapsulation, already surpasses previously reported colloidal nanocrystal lasers, which typically fail within seconds to a few minutes under similar conditions.[37, 64]

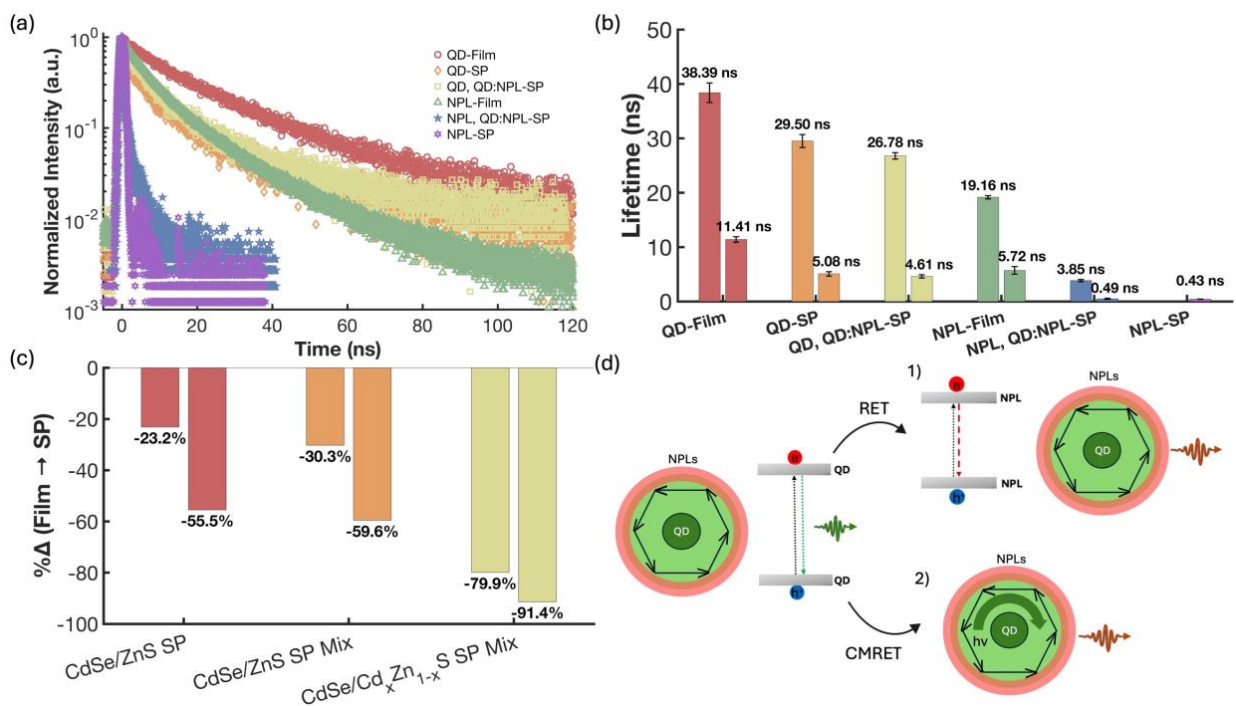

Figure 4: Time-resolved photoluminescence (TRPL) analysis of QD and NPL systems in film and SP architectures. (a) Normalized PL decay curves for QD and NPL films and their corresponding SPs and (b) their extracted biexponential lifetimes. (c) Percent reduction in lifetimes from films to SPs. (d) Proposed energy-transfer mechanisms in the spherical SP. In the primary pathway (1), QDs radiatively excite NPLs directly, with NPLs positioned near the SP boundary, overlapping spatially with whispering gallery mode (WGM) maxima. In the alternative, secondary pathway (2), QD emission couples first into WGMs, and subsequently these modes radiatively excite the boundary-assembled NPLs.

The long-term spectral stability demonstrated in Figure 3 suggests efficient and resilient carrier dynamics within the SPs. To probe the microscopic origin of this stability, we performed time resolved photoluminescence (TRPL) measurements (Figure 4) to characterize exciton lifetime changes upon assembly of QDs and NPLs into SPs. Decay curves (Fig. 4a) are fit (>97% confidence, except IRF limited fits, See SI Table 8.1) to either single or biexponential functions (Fig. 4b), yielding a longer decay component ($\tau_1$) and a shorter decay component ($\tau_2$). These lifetime reductions are quantitatively summarized in Figure 4(c). In general, the assembly of QDs or NPL into SPs triggers a reduction in lifetime as compared thin films. Since thin films of QDs or NPLs retain similar interparticle interactions as to SPs but lack strong cavity effects, we attribute the reduced lifetimes to enhanced recombination mediated by cavity effects. Two internal checks disfavor a purely local-field origin: under identical optics, NPL-only SPs do not lase, whereas hybrids do; moreover, the narrow emission peaks exhibit free-spectral-range spacings that scale with the measured SP diameter, as expected for true WGMs.

Specifically, the QD-SP (orange open diamond, and bar) exhibits lifetimes of $\tau_1$= 29.50 ns (23.2% reduction) and $\tau_2$= 5.08 ns (55.5% reduction) compared to QD-film ($\tau_1$= 38.39 ns, $\tau_2$= 11.41 ns).

When QDs are incorporated into hybrid QD-NPL SPs, they exhibit slightly shorter lifetimes (gold-yellow open squares and bar) of $\tau_1$ = 26.78 ns (30.3% reduction) and $\tau_2$ = 4.61 ns (59.6 % reduction) confirming the presence of additional decay pathways due to at least partial energy transfer to nearby NPLs.

The NPL-SP (purple open hex) exhibits a single exponential decay with a short lifetime of $\tau$ = 0.43 ns, the fastest recorded in our system and partially convoluted within the instrument response (~400 ps), precluding a meaningful biexponential fit. This decay is significantly shorter than the reference NPL thin film ($\tau_1$ = 19.16 ns, $\tau_2$ = 5.72 ns), confirming the crucial role of cavity effects. NPLs assembled within the hybrid SP (blue open star and bar) also display dramatically short lifetimes of $\tau_1$ = 3.85 ns and $\tau_2$ = 0.49 ns, corresponding to substantial rate enhancements relative to the NPL-film ($\tau_1$ reduced by 79.9%, $\tau_2$ by 91.4%). Notably, the hybrid NPL lifetimes are slightly longer and better described by a biexponential model compared to the NPL-SP, possibly suggesting partial transient replenishment of exciton populations via energy transfer from QDs, consistent with the donor-to-acceptor transfer evidenced by QD lifetime shortening. The strongest acceleration appears in the NPLs located at the periphery, consistent with WGM field maxima at the boundary, and is inconsistent with a uniform local-field enhancement or volume-wide defect localization.

We propose that the moderate reduction of the QDs lifetime and dramatic shortening of NPLs lifetime when assembled within the hybrid SPs support a core-shell spatial arrangement, where QDs predominantly occupy the core region of the SP and NPLs form an outer shell. We expect such architecture to limit Förster resonance energy transfer (FRET), as only interfacial QDs lie within the critical Förster radius for significant transfer efficiency. We estimate the FRET efficiency using a modeled Förster radius of ~ 4.5 nm, based on previously reported values ($\mathcal{R}_0 \approx$ 6 − 9 nm).[65-68] For efficient transfer, QD-NPL separation would need to be ≤ 4.5 nm. However, STEM measurements indicate a best-case spacing of ~9 nm, nearly double that distance. This calculation, detailed in the supporting information, yields a maximum FRET efficiency of only ~ 1.5%. Consistent with this interpretation, Figure S3 further compares the relative QD and NPL emission intensities, showing that their ratios follow the nominal composition rather than the strong quenching/enhancement expected for efficient FRET. Because the SP spectra in Figure S3 were normalized to the maximum emission intensity, these trends are intended to be interpreted qualitatively rather than as quantitative compositional ratios. Such negligible transfer supports that the observed NPL lifetime reductions primarily result from photonic confinement effects, such as the Purcell effect, rather than non-radiative QD-to-NPL energy transfer. Together with the absence of sub-band trap emission and the type-I CdSe/ZnS alignment that confines carriers to the QD core, interfacial-trap or FRET-dominated pathways cannot account for the observed kinetics or lasing.

We propose a photonic mechanism depicted in Figure 4(d). Pathway (1, RET): QDs radiatively transfer energy to nearby NPLs via FRET, after which the NPLs emit photons that couple into WGMs at the SP boundary. Pathway (2, CMRET): QDs radiatively emit photons that first couple into WGMs, and these cavity fields subsequently excite NPLs at the periphery. Considering the measured QD-to-NPL separation (~9 nm), the negligible calculated FRET efficiency (~1.5%), and the pronounced lifetime shortening of NPLs, we rule out Pathway 1 as a dominant mechanism. Instead, the evidence supports Pathway 2, where cavity-mediated radiative excitation enhanced by

WGMs primarily drives rapid recombination and selective lasing behavior in the NPLs. This mechanism highlights the cooperative optical roles of QDs as broadband absorbers and NPLs as high-gain emitters, jointly sustaining stable lasing through cavity-enhanced energy transfer.

Conclusions

We have introduced a colloidal platform that integrates inter-NC radiative energy transfer with cavity enhanced light matter interactions in a single, self-assembled architecture. Hybrid SPs composed of CdSe/ZnS QDs and CdSe/Cd$_x$Zn$_{1-x}$S NPLs exhibit low-threshold lasing (0.35 mJ/cm$^2$) with minimal spectral drift (<0.8 nm) and strong photostability, maintaining consistent emission under prolonged excitation and high fluences (1 mJ/cm$^2$ for 20 min) in air. This performance results from efficient cavity-mediated energy funneling from QDs to NPLs and photonic confinement, addressing key limitations in colloidal lasing systems, including spectral instability and environmental sensitivity. Notably, the architecture is substrate free, solution processable, and optically durable, making it well suited for scalable applications beyond microlasers. These include functional pigments with tunable and stable emission, photonic inks, and optical security features, where robustness and spectral precision are critical. Future improvements in inorganic shell and organic ligand design could extend the operational fluence range by mitigating Auger recombination, therefore enhancing flexibility for continuous or higher-power operation.

Methods and Experimental

Synthesis of CdSe Nanoplatelets (NPLs)

Four-monolayer CdSe nanoplatelets were synthesized using a modified hot-injection procedure based on established literature methods.[69] Cadmium myristate, selenium, and 1-octadecene were combined under inert atmosphere, degassed, and heated to induce 2D platelet growth, followed by controlled quenching and purification to selectively remove 3-monolayer platelets. NPL thickness was confirmed by absorption spectroscopy. Full experimental details are provided in the Supporting Information.

Synthesis of CdSe/Cd$_x$Zn$_{1-x}$S Core/Shell NPLs

CdSe/Cd$_x$Zn$_{1-x}$S core/shell NPLs were prepared by growing an alloyed Cd$_x$Zn$_{1-x}$S shell onto 4-ML CdSe platelets using cadmium and zinc oleate precursors. A slow, controlled injection of octanethiol in ODE was used to regulate shell growth, with precursor ratios and reaction duration tuned to achieve uniform alloy composition and shell thickness, following adaptations of reported procedures.[69] Full experimental details are provided in the Supporting Information.

Synthesis of CdSe/ZnS Quantum Dots

CdSe/ZnS core/shell quantum dots were synthesized by injecting TOPSe into a Cd(oleate)$_2$ precursor solution in ODE and subsequently depositing CdS and ZnS shells using thiol and phosphine-sulfide precursors, respectively, following modified literature methods.[70] The nanocrystals were purified through repeated precipitation and redispersion and stabilized with oleic acid. Full experimental details are provided in the Supporting Information.

Preparation of Hybrid QD:NPL Supraparticles

Hybrid supraparticles were fabricated using a microfluidic source–sink emulsion approach adapted from prior supraparticle assembly work.[45] A 1:1 (v/v) mixture of CdSe/ZnS QDs and CdSe/Cd$_x$Zn$_{1-x}$S NPLs in toluene was emulsified into SDS-containing water, where gradual solvent extraction drove droplet densification. Shape and surface-energy differences caused NPLs to enrich the droplet periphery while QDs concentrated toward the interior, producing radially graded supraparticles. Full experimental details are provided in the Supporting Information.

Optical Characterization

Steady-state and fluence-dependent photoluminescence measurements were performed on a home-built, fiber-coupled confocal microscope. Excitation at 488 nm from an OPA–Monaco laser system was filtered, fiber-delivered, and focused through a 50× 0.8 NA objective to a ~12 µm spot. Emission was collected through the same objective, filtered, and directed to a Horiba iHR550 spectrometer equipped with 1200 g mm$^{-1}$ or 150 g mm$^{-1}$ gratings for high-resolution WGM analysis or broadband spectra, respectively. Absolute PL quantum yields were measured using an Edinburgh FLS1000 integrating sphere. Time-resolved PL was recorded using TCSPC and fit with single- or bi-exponential models as appropriate. Power values refer to the measured power at the sample plane. Full experimental details are provided in the Supporting Information.

Electron Microscopy

SEM images were acquired on a Tescan S8252X operated at 5 keV under ultra-high-resolution scan mode. High-resolution TEM and STEM imaging were performed on a JEOL-F200 at 200 kV. Sample preparation procedures were optimized to minimize partial supraparticle disassembly during drying. Full experimental details are provided in the Supporting Information.

Whispering-Gallery Mode (WGM) Analysis

Lasing peaks were assigned to whispering-gallery modes by comparing measured emission wavelengths with solutions to the approximate spherical resonator condition using an effective refractive index of ~1.83, following standard WGM treatment.[71] TE/TM splitting was evaluated using polarization-dependent correction terms, and representative field profiles were validated by FDTD simulations. Full experimental details are provided in the Supporting Information.

FRET and Purcell Factor Estimation

Förster energy-transfer efficiencies between QDs and NPLs were estimated using donor quantum yield, spectral overlap, and morphology-derived donor–acceptor distances. Purcell enhancement factors were extracted from QD and NPL lifetime shortening in supraparticles relative to films under the assumption of unchanged nonradiative decay rates. Full experimental details are provided in the Supporting Information.

Supporting Information

Detailed procedures for the synthesis and purification of CdSe/ZnS quantum dots and CdSe/Cd$_x$Zn$_{1-x}$S nanoplatelets; preparation and characterization of hybrid QD–NPL supraparticles; additional steady-state and excitation-dependent PL measurements; fluence-dependent lasing data and extraction of peak positions, FWHM values, and Q-factors; SEM and TEM micrographs of supraparticle morphology; extended lifetime fitting and analysis of donor–acceptor energy transfer; supporting derivations for WGM mode assignments and effective-index fitting; and supplemental figures and control experiments.

Acknowledgments

The supraparticle synthesis and optical measurements were supported by the NSF STC-IMOD under award DMR-2019444. The optical modeling was supported by the NSF ERC IoT4Ag under grant EEC-1941529. Electron microscopy was carried out at the Singh Center for Nanotechnology, which is supported by the NSF National Nanotechnology Coordinated Infrastructure Program under grant NNCI-2025608. C.G. acknowledges support from the NSF Graduate Research Fellowship Program (NSF-GRFP). C.Y.K. acknowledges support from the Vagelos Integrated Program in Energy Research (VIPER). E.M. acknowledges funding from the European Research Council through ERC Starting Grant REFINE2LASE (grant agreement n. 101202342)).